\documentclass[conference]{IEEEtran}
\IEEEoverridecommandlockouts
\usepackage[utf8]{inputenc}
\usepackage{cite}
\usepackage{amsmath,amssymb,amsfonts}
\usepackage{algorithmic}
\usepackage{graphicx}
\usepackage{textcomp}
\usepackage{xcolor}

\usepackage{latexsym}
\usepackage{mathptmx}
\usepackage{amsbsy}
\usepackage{amsthm}
\usepackage{array}
\usepackage{multirow}
\usepackage{hhline}		
\usepackage{subfigure}
\usepackage{mathptmx}
\usepackage{mathtools}
\usepackage{todonotes}

\usepackage{enumitem} 
\usepackage[square,sort&compress,comma,numbers]{natbib}

\usepackage{booktabs} 
\usepackage{float}
\usepackage{scrhack}
\usepackage{listings}
\usepackage{color}
\usepackage{pgfplots}
\usepackage{tikz}
\usepackage{cleveref}
\usepackage{here}
\usepackage{url}
\usepackage{enumitem}
\usepackage{lipsum}
\usepackage{eqnarray}

\usepackage{breakurl}

\def\BibTeX{{\rm B\kern-.05em{\sc i\kern-.025em b}\kern-.08em
    T\kern-.1667em\lower.7ex\hbox{E}\kern-.125emX}}
\begin{document}

\title{Automated Enterprise Architecture Model Mining}

\author{\IEEEauthorblockN{Peter Hillmann, Erik Heiland, and Andreas Karcher}
\textit{Universität der Bundeswehr München}\\
Werner-Heisenberg-Weg 39, 85577 Neubiberg, Germany \\
Email: \{peter.hillmann, erik.heiland, andreas.karcher\}@unibw.de}

\maketitle

\begin{abstract} 
Metadata are like the steam engine of the 21st century, driving businesses and offer multiple enhancements.
Nevertheless, many companies are unaware that these data can be used efficiently to improve their own operation.
This is where the Enterprise Architecture Framework comes in.
It empowers an organisation to get a clear view of their business, application, technical and physical layer.
This modelling approach is an established method for organizations to take a deeper look into their structure and processes.
The development of such models requires a great deal of effort, is carried out manually by interviewing stakeholders and requires continuous maintenance.
Our new approach enables the automated mining of Enterprise Architecture models.
The system uses common technologies to collect the metadata based on network traffic, log files and other information in an organisation.
Based on this, the new approach generates EA models with the desired views points.
Furthermore, a rule and knowledge-based reasoning is used to obtain a holistic overview.
This offers a strategic decision support from business structure over process design up to planning the appropriate support technology.
Therefore, it forms the base for organisations to act in an agile way.
The modelling can be performed in different modelling languages, including ArchiMate and the Nato Architecture Framework (NAF).
The designed approach is already evaluated on a small company with multiple services and an infrastructure with several nodes.

\end{abstract}


\section{Motivation for Enterprise Architecture}
\begin{figure}[bh]
	\vspace*{-0.2cm}
	\centering
	\includegraphics[width=0.49\textwidth]{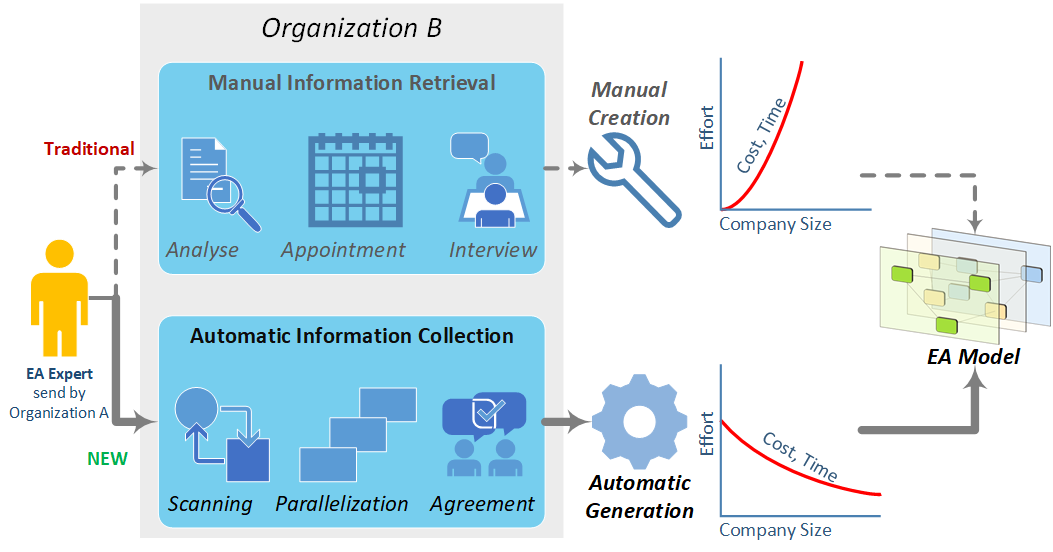}
	\caption{Typical situation of an enterprise architect expert.}
	\label{fig:problem}
	\vspace*{-0.1cm}
\end{figure}

Mastering IT landscapes that have grown over many years with their often complex structures requires a systematic approach for further development.
Due to the ever shorter innovation cycles, the alignment of the business and IT need to be sufficient flexible.
In addition, processes must be adapted to new conditions and should be optimized continuously.
Especially, the adaptation of these in the course of digitalization requires a high degree of transparency and scalability.
It is unreasonable for an existing company to reinvent itself completely.
Incremental changes are in most cases a more resource efficient tactic.
The established supporting toolkit for this problem is Enterprise Architecture.
It provides detailed views form different hierarchy levels and several perspectives.
These range from top-level corporate management to IT administration at the lowest level.
Established views are business, application and technical, whereas these are connected with each other.
It thus forms the basis for a overarching communication and provides a continuous control mechanism.
In the end, it leads to optimized processes and improved business with reduced costs.
The problem lies in the high effort of manually creating the corresponding documentation.
At present, Enterprise Architecture models are created by hand.
It is carried out manually by interviewing stakeholders and requires continuous maintenance.
Furthermore, this usually disturbs the employees during their work.
This leads often to obsolete models and missing details.
These do not reflect the current status when the IT landscape changes.
Fortunately, there is a lot of data available within a company from various sources that can be used to generate such models.
The typical challenge with the current manual and the new automatic generation approach is visualized in Figure \ref{fig:problem}.

The objective of our approach is the automated creation of Enterprise Architecture models based on captured information from a company's existing IT infrastructure.
The data mining is done autonomously, using different sources such as network data, log files and process documentation.
These data are used for the automated creation and maintenance of enterprise architecture models representing the companies business.
The collection of IT-relevant information and process-related activities not only leads to enormous time savings, it also ensures a higher quality of the resulting models.
The advantage lies in the evaluation of live operating data that has not already been misinterpreted or aggregated.
This can be the case because enterprise architects are usually not involved in the technical aspects of the projects.
Therefore, they rely on the input from the individual departments.
As a result, these models are created which faithfully depict the current company events in detail, enabling targeted management, controlling, and optimization of the business IT and processes.
The generated models can be mapped to different modelling languages or frameworks.
ArchiMate and NATO Architecture Framework (NAF) are mentioned here as two most common targeted enterprise architecture frameworks of more than 50~\cite{Matthes2011}.



\section{Scenario and requirements}\label{sec:scenario}
The challenge is based on the following problem statement.
A company specializes in the management of supply chains and offers various digital services for this purpose.
The enterprise has gone through several stages of development from a start-up to a small and medium-sized enterprise (SME).
It has grown over the past years and complex structures have evolved as a result.
In order to reduce costs and improve competitiveness, processes have to be optimized and IT needs to be better aligned with the business.
Among other things, a higher resilience against cyber attacks is to be achieved.
Currently, we are expecting no established EA Management at all, which means missing documentation or central descriptions of the different layers and views of the organisation.
Therefore, the company hires an external expert, to establish a controlling workflow and to establish EA Management for the company and their business.
In order to be able to systematically record and analyse these infrastructures, EAM has already established itself as a discipline in many organizations.
However, the modelling and maintenance of such architectures by hand is a great effort.
Currently, it is done manually by reading documents, analysing data and interviewing employees.
On the one hand, the involved actors often are not able to create such an architecture for a corresponding system design. 
On the other hand, the enterprise architects, who have mastered the modelling, lack the knowledge about the systems.
In both cases, this discrepancy leads to an inadequate representation of reality when it comes to describing the underlying systems and processes.
Decision-makers who rely on these models, may then act incorrectly when it comes to procuring new IT or adapting business structures.

The management of enterprise architectures is an iterative process and aims to improve a business-IT alignment.
Rapid technological change as well as ever new cyber threats from outside require a constant review and adaptation of the underlying IT infrastructure of an enterprise~\cite{Luftman2000AssessingBA}.

A model-based depiction of the actual situation using automation mechanisms saves a lot of time, ensures the quality of the models and leads to a better controllability of the IT landscape.
This highly supports the defence possibilities against cyber threats.

However, the data are almost always heterogeneous and usually represent only fragmentary views of certain aspects.
In order to bring such data together and obtain a uniform view of the company, a suitable methodology is required.


Based on this, we obtain the following requirements~\cite{10.1007/978-3-642-34163-2_2}:
\begin{itemize}
\item \textbf{Data Extraction}:\\ Collection of metadata from different sources and their harmonization and correlation.
\item \textbf{Integration}:\\ Heterogeneous information sources have to be integrated and mapping to processes to form a knowledge base.
\item \textbf{Modelling}:\\ Selective mapping between collected data and EA model.
\item \textbf{Manual Enrichment}:\\ Adding additional information based on manual side information.
\end{itemize}

The characteristic of this scenario is similar to a smart data problem.
The heterogeneity of the different sources and metadata structures creates in relation to several companies a large challenge.
In these sources, recurring structures must be extracted from large amounts of data using coordinated algorithms.
Here, specific approaches from the field of AI can also be used.

Therefore, the following research questions will be addressed:
\begin{enumerate}
\item Which Enterprise Architecture Framework is suitable for automated model generation, presentation and validation?
\item What can be achieved with an automated approach in the context of EA?
\item To what extent can the different levels of abstraction and entity relationships between them be automatically represented in the model?
\item How can the system be permanently integrated into the organization without any disabilities or security hazards?
\end{enumerate}

\section{Related Work}
This section is divided in two parts: approaches for EA modelling and technologies for automated data mining.

An EA model is highly affected on the changes in IT, which is usually driven by the requirements engineering.
The EA management performs to document and analyse the current EA of an organization to plan the digital transformation~\cite{mci/Aier2008}.
The diversity of over 50 EA frameworks offers a lot possibilities in management~\cite{Matthes2011}.
Here, we focus on a widely used frameworks, which are close to our requirements.
One of the newer modelling languages is ArchiMate~\cite{OpenGroup2017}, which was published in the year 2004.
It is strongly related to TOGAF~\cite{OpenGroup2018} and is based partially on IEEE 1471 standard~\cite{Maier2000}.
The main advantage is the simplicity.
The most established framework in the field of IT is ITIL~\cite{10.5555/2222671,RePEc:zbw:simata:0412019}.
With its approach of customer orientation and the experience gained from best practice, it is considered internationally as a de facto standard without auditability.
That's why, ITIL should be supplemented by ISO/IEC 20000~\cite{10.5555/2331268} for IT service management and ISO 27000~\cite{Disterer2013} for IT security.
A further, supplementary framework is FitSM~\cite{Froeschle2019}, which offers a multiple checklists for a quick documentation.
Last but not least, the military approach for EA is the NATO Architecture Framework (NAF)~\cite{NATOACT2018}.
It also offers a metamodel for the description of an enterprise.
As we focus on the realization of things and processes, frameworks on the higher meta-level like IT4IT, COBIT and Six Sigma are out of scope.
Despite the many frameworks, none of them offers a suggestion to automate the modelling process.
Nevertheless, one of the commonalities of the frameworks is the division into viewing levels, which influences on our concept.
A usual distinction is made between the business view, application view, and technology layer.
Furthermore, a separation is made between passive and active structures.

\begin{figure*}[t]
	\centering
	\includegraphics[width=1.0\textwidth]{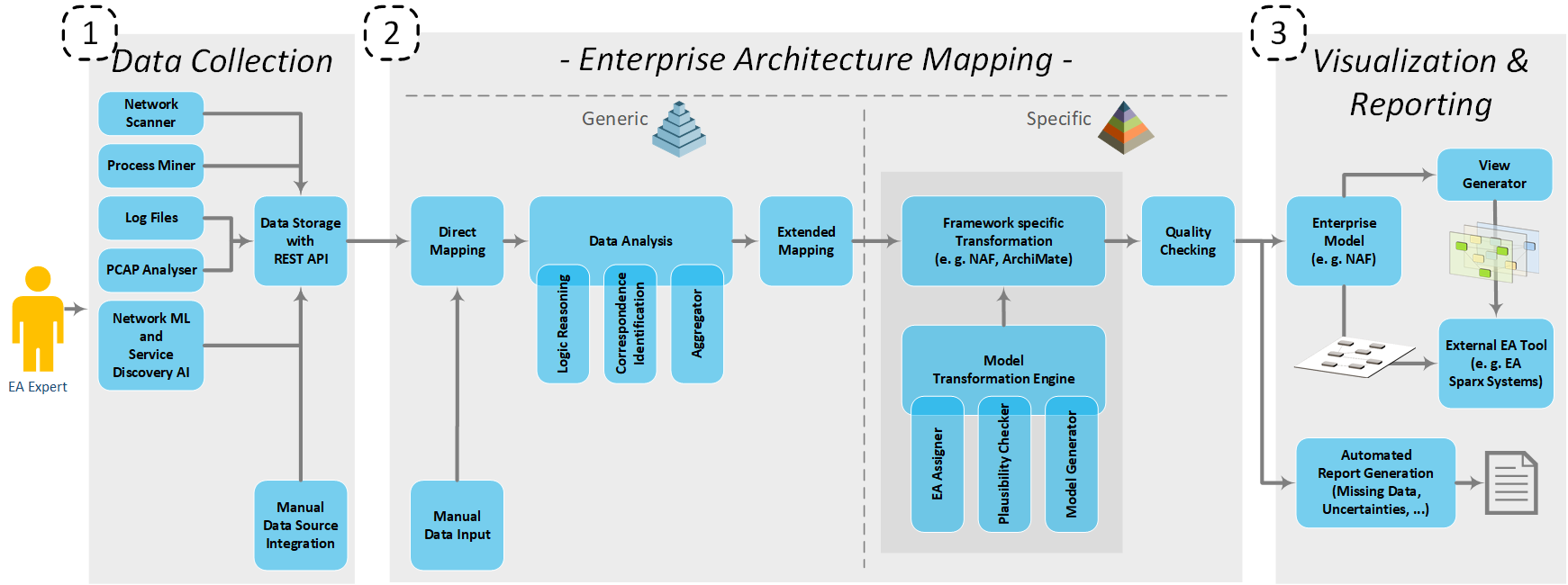}
	\caption{Overview of the process of the automated enterprise architecture modelling approach.}
	\label{fig:AutoEA_overview}
	\vspace*{-0.2cm}
\end{figure*}

To perform automated data collection, several techniques and methods should be considered~\cite{Gort2007}.
In the context of infrastructure monitoring, tools like Nagios~\cite{NagiosEnterprises2020} and Wireshark~\cite{10.5555/1202316} or protocols such as SNMP~\cite{10.17487/RFC1157} are usually used.
Furthermore, methods of network scanning as NMAP~\cite{10.5555/1538595} provide basic information about the IT landscape.
To further extend, the network topology can be visualized with graphs for example by Icinga~\cite{IcingaGmbH2009}. 
It also includes first steps for process descriptions.
Beside this, approaches of \textit{Security Information and Event Management} (SIEM) and \textit{Open Source Intelligence} (OSINT) present already aggregated and consolidated information of the technology layer of an enterprise.
These information can be used for our approach.
Other sources are log files, which can be collected by a log-server like Flume~\cite{Nagdive_2019} and Rsyslog~\cite{AdisconGmbH2020}
The solutions of Docusnap~\cite{GmbH2020} and LANSurveyor~\cite{SolarWindsWorldwide2020} have a lot features to visualize, document and analyse the IT infrastructure.
In the end, all technical approaches offer a lot capabilities for automatic data collection.
Nevertheless, these tools fail in terms of the model-based description with regard to enterprise architecture.
The idea of a service orientation or capability driven approach is not recognized.

All in all, the presented techniques for automated description do not go beyond the technical aspects.
Especially, the higher layers and the more abstract views for business and the horizon of manager are missing.
Also the automated discovery and description of processes chains are new aspects.
In addition, current possibilities lack in flexible modelling in different EA frameworks.
In the end, none of the examined possibilities provides a solution for the problem.

\section{Concept of Model Mining}
Based on the available data sources, our concept for \textit{automated EA model mining} is in general a bottom-up approach.
It enables the enterprise architect to create a snapshot of the IT landscape as required.
On this basis, our approach further pursues the goal of a holistic picture of the company according to the specification of EA management.
The general process can be divided in three steps. 
First of all, the \textit{data collection} builds the data storage for the further steps.
All collected data from the various heterogeneous sources are stored in a harmonized way.
In the second step, the \textit{EA mapping} takes place.
This process is split in two sub-parts.
In the first part, a model is created by assigning the analysed data elements according to a generic model description.
Afterwards, the uniform model elements are then mapped to the desired EA target metamodel.
The last step is responsible for the \textit{visualization} and \textit{reporting}.
Thereby, the created model is prepared for further processing in other EA tools as well as a presentation to obtain an overviews.
Beside this, a report is generated to align the result with guidelines according to the common EA framework's.
The entire process is presented with the main components in Figure~\ref{fig:AutoEA_overview}.


The application within a company initially requires close cooperation between the enterprise architects and the responsible IT departments.
It must be ensured that the system landscape to be analysed is in operation at the time of data collection.
If necessary, appropriate rights must also be granted in order to be able to perform the corresponding analyses.

\subsection{Data Collection}
The objective of this phase is to set up a central instance that holds all data necessary for the specific purpose.
At the beginning, an analysis of the available data sources is carried out in the viewed company.
These are then evaluated with regard to their relevance for our approach.
The sources considered useful are then connected to the central data storage, which serves as a data sink from this side.
Preferably, this is done via a read-only \textit{Representational state transfer} (REST) interface at a specific source.
As the data storage can contain sensible information, it needs to be secured properly against IT security attacks.
To match these requirement, the developed approach can be implemented inside the company by themself.
So, the sensitive information remains at the company's own and controlled environment.
For an automatic monitoring of the core layers of an EA model is more than one data input source mandatory~\cite{Kleehaus2017}.
As primarily source serves network scanner and system analysis tools to automatically collect and identify elements in the IT landscape.
In addition, typical useful sources for the use case are documentation, log files, network traffic, process descriptions and existing EA models.
This refers to overall hierarchically levels from the technology to business.
It includes operating system, applications and services, whether it belongs to an internal and external connected processes or partner.
Furthermore, the extension with additional information sources is possible preferably via a REST-API.
The automated collection of this data is comparable to the scanning phase of penetration tests~\cite{8945857} or comes close to a forensic investigation~\cite{schopp2020agile}.

\begin{figure}[bht]
	\centering
	\includegraphics[width=0.49\textwidth]{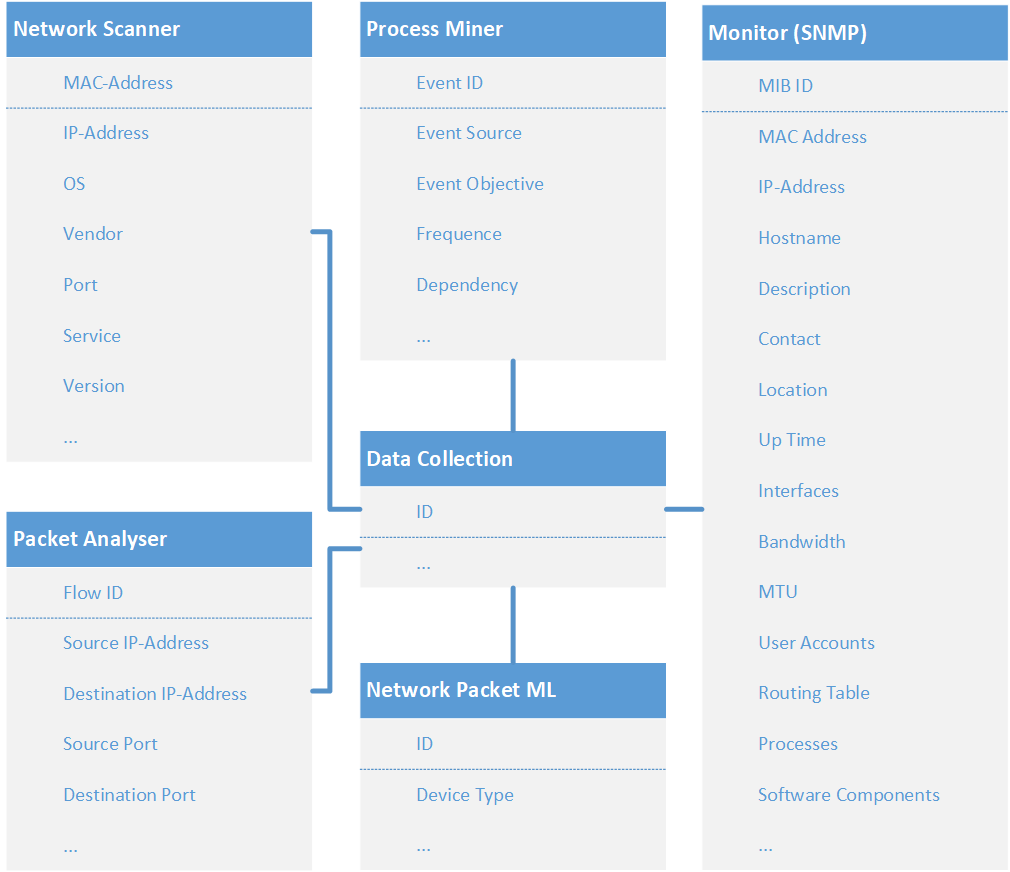}
	\caption{Extract from the possible data sources.}
	\label{fig:datasources}
	\vspace*{-0.2cm}
\end{figure}

The difficulty lies in the heterogeneity of the different data sources.
In addition, several companies may or may not have specific sources available.
All identified information are harmonized in relation to format and data structure.
Especially, the elements of the metadata are of interest.
These provide an insight into what is happening in the company without causing data protection problems.
An excerpt of these data model is showed in Figure~\ref{fig:datasources}.
We decided on a simple list format to easily and flexible extend the approach to other environments or EA frameworks.

\subsection{Enterprise Architecture Mapping}
The second phase controls the creation of an EA model.
First, the identified data points are mapped to a generic intermediate format.
Then the transformation takes place with regard to a specific EA framework.
This two step approach offers more flexibility.

\paragraph{Generic Model Preparation}
In the first subphase, a generic model preparation takes place.
This temporary intermediate step enables free modelling for an enterprise architecture model without the specifications of a specific framework.
The generic description also forms the basis for the subsequent flexible mapping to various specific frameworks.
Therefore, we created a architecture catalogue, which contains the unification of the elements and relations of value.
According to our uniform data format, it is constructed as a list of elements and relations.
So this approach is easily extensible in the future.

Based on the raw data, a direct mapping is performed of the identified data points to the metamodel elements of the generic modelling.
This static description follows a simple 1-to-1 rule-based approach. 
It contains a comprehensive collection of entities, relations and attributes. 
During this process, manual data input is also possible.
Table \ref{tab:datacollection} presents an extract of these rule set.
It contains the metadata element, the description, and the current automation degree.

\begin{table}[hbt]
	\centering
	\caption{Excerpt of the main data elements for EA modelling.}
	\begin{tabular}{lll}
		Generic Element        	 & Data Source                  & Automation \\
		\hline
		Node                     & IP, MAC address              & auto              \\
		Device                   & MAC address, device type 	& auto              \\
		System Software          & Ports, Operating System (OS) & auto              \\
		Application Interface    & Port                 		& auto              \\
		Application Collaboration & n/a                         & manual            \\
		IT Service 				 & Port, Log Files              & semi-auto         \\
		Path                     & Src./dst. Address            & auto              \\
		Communication Network    & IP subnet                    & auto              \\
		Technology Function      & Process, event               & semi-auto         \\
		Technology Process       & Process, Log Files           & semi-auto         \\
		Technology Interaction   & Event                        & manual            \\
		Technology Event         & Event Log                    & semi-auto         \\ 
		Business Actor / Role	 & OS Accounts, SNMP			& semi-auto			\\
	\end{tabular}
	\label{tab:datacollection}
\end{table}

This is followed by the analysis of the data.
Here, defined rules and logical conclusions are used to further process the information.
By aggregating the data from different analysis tools, a holistic representation of the underlying system landscape is created in this phase.
It connects multiple elements with each other and deducts relationships.
At this point, it identifies missing elements according to logical IT services structures and processes.
So, it enriches the information base with additional elements, called dummy objects.
These dummy objects does not exists in the first data collection phase, but these should exist for example in context of an EA.
Later on, these dummy objects are listed in the final report for verification.
The data analysis also identifies duplications.
The process of information mining is described similar to higher-order logic.
Furthermore, the extracted events in the raw data can be analysed by process mining to obtain event steps of a value chain.
At this point, also machine learning is possible to determine hidden elements.

For example, a network scanner collects information about IP address, port, and MAC address.
From the IP address is derived a device on the technology layer.
To classify the device with a type, the network packet machine learning technique is used.
Furthermore, the open ports allow estimating the operating system and their version as well as its hosted services.
From these information, an application component is reasoned.
Usually, a service or application is generating log files that the EA architect can extract at the machine.
With an event log, the process mining technique reconstructs a business process.
In addition, further information can be queried via \textit{Simple Network Management Protocol} (SNMP) services from the \textit{Management Information Base} (MIB) and \textit{Object Identifier} (OID).
This allows the modelling of a business actor as well as further EA elements and relationships.

\begin{figure}[t]
	\centering
	\includegraphics[width=0.49\textwidth]{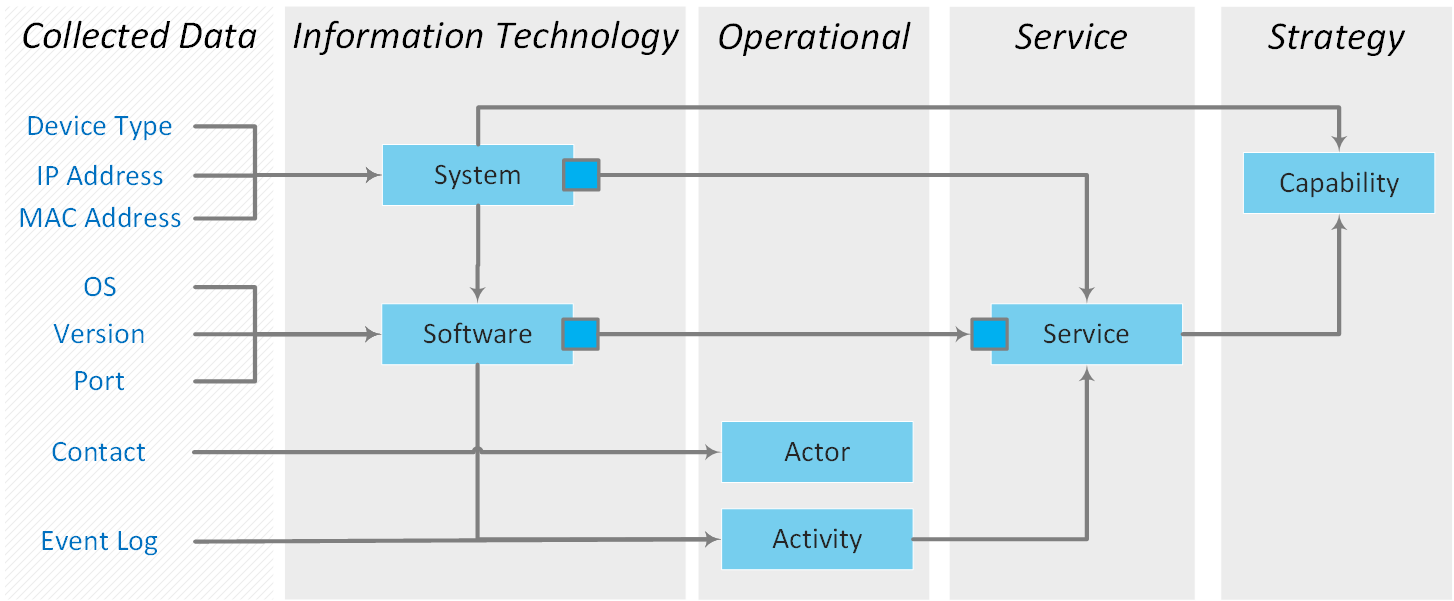}
	\caption{Example of model generation and reasoning for a universal EA framework.}
	\label{fig:AutoEA_detailled}
	\vspace*{-0.3cm}
\end{figure}

The process of reasoning is visualized in Figure~\ref{fig:AutoEA_detailled}.
Based on the collected and aggregated data this allows to draw conclusions about further EA elements.
With this inside-out approach one obtains model components on both other views, higher and lower layers.
In particular, views on the business level can be generated from the automatically determined information on the technology level and application level.
Thus, the pursued goal of a better control possibility for the management is achieved by providing condensed information without technical details.


Subsequently, the extended mapping process transfers the new data points to model elements within the generic framework.
Finally, the relationships between all model elements are inserted in the model.

\paragraph{Specific Model Transformation}
The second subphase is to map the generated model to a specific EA framework.
Using a transformation engine developed specifically for such purposes, the model-based formulation of rules allows the information to be mapped to any target model~\cite{Heiland2019}.
This decoupling step is advantageous if the organization decides to switch to a different framework or even a new version.
Similarly, if a heterogeneous EA landscape exists using multiple EA tools, the automated approach can still be used.

Nearly all frameworks have in common the three layered structure with business, application, and technology with pyhsical elements.
Also the categorization of the elements in passive, active and behaviour is recognizable.
Nevertheless, the main challenge is the ambiguity of the translation of information to the specific metamodel of a framework~\cite{Holm2014,6037637}.

The difficulty in this step is the definition of the individual elements by a specific EA framework and their relationships.
In addition, architecture building blocks are often defined for specific contexts.
For example, the Framework IT4IT describes a value stream always with the following components according to the reference architecture~\cite{OpenGroup2017a}: \textit{Functional Component}, \textit{System of Record Integration}, \textit{Lifecycle Data Object}, \textit{Service Backbone Data Object}, \textit{Relationship}, and \textit{Multiplicity}.
In contrast, the Framework ArchiMate is structured based on services.
The modelling pattern for an application~\cite{OpenGroup2017} consists of: \textit{Data Object}, \textit{Service}, \textit{Function}, \textit{Interaction}, \textit{Interface}, \textit{Component}, and \textit{Collaboration}.
Moreover, the framework ITIL has a more complex metamodel of an IT service~\cite{Braun2007IntegrationOI}.

To take this into account, a \textit{Model Transformation Engine} is developed.
It consists of the \textit{EA assigner}, which maps the generic elements to the specific elements of the target framework.
In the \textit{plausibility checker}, the specifications of the metamodels and the \textit{architecture building blocks} are inspected.
A suitable transfer to the model is calculated.
The \textit{model generator} then generates the adapted model.
In addition, information about the model changes is generated for later reporting.

\subsection{Visualization and Reporting}
The final step includes the visualization and documentation of the generated output.
Depending on the framework used, there are different possibilities for the views to be generated.
Further requirements arise in the choice of modelling tool and description language.
For the combination of these factors, a view generator is finally implemented to meet the requirements of an EA model and to further reduce the modelling effort.
According to universal output format, XML is chosen.

The automatically generated report provides the architect with additional information necessary for post-processing the models.
It shows, for example, where information is still missing or if there are uncertainties whether a service will run on the assigned hardware.
According to frameworks like ITIL, IT4IT and FitSM, the reporting is enriched with the provided reference architectures and architectural building blocks for a specific case.
At this point, the exchange between the architect and the IT experts or the management can begin in order to agree on details and future target architecture.

\section{Evaluation}
\begin{figure}[hbt]
	\vspace*{-0.2cm}
	\centering
	\includegraphics[width=0.49\textwidth]{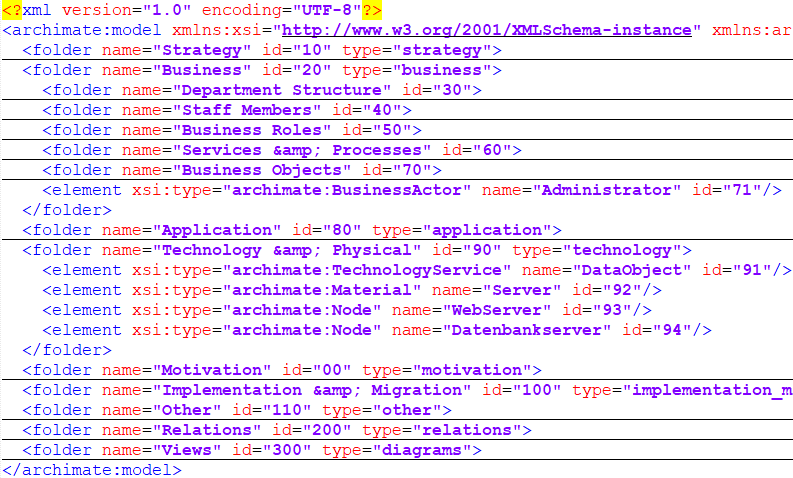}
	\caption{Snippet of XML file from tool Archi.}
	\label{fig:Listing3a}
\end{figure}

For the feasibility study and potential analyses, the concept focus on a mapping to the EA modelling language of ArchiMate.
The documentation is freely available, it is easy to use, and the open source modelling tool Archi provide visualization to an EA model.
The tool Archi uses to import and export EA model files based on the ArchiMate Exchange Format~\cite{OpenGroup2017} (similar to XML), so it is the output of our prototype.
Figure~\ref{fig:Listing3a} shows a snippet of an Archi XML file.
%

We analysed every element of the core framework to obtain a clear assignment between the data collection models and the target metamodel.
At the bottom of the data collection stack is the technology layer located.
The elements are used to model the IT architecture of the organisation.
It describes the structure and the behaviour of the technology infrastructure with its physical components and some implementation details.
The next layer is the application layer on top of our data collection stack and belongs to the core framework.
The application layer elements are used to model the application architecture of the organisation.
It describes the structure, the behaviour and the processes of the applications of the enterprise.
The business layer elements are on the top of the application and data collection stack perspective.
It is modelling the operational organisation of an enterprise and their external connections.
The model is technology independent and provides the strategy as well as the vision for the future.
In addition, the ArchiMate language provides a set of generic relationships for the universal elements.
For the current concept, the simple association relationship is the default value for all relationships of the automatic EA model mining.
For a more precise description, the EA architect has to specify the relationship with his experience and knowledge after the creation of the model.

\begin{figure}[hbt]
	\centering
	\includegraphics[width=0.49\textwidth]{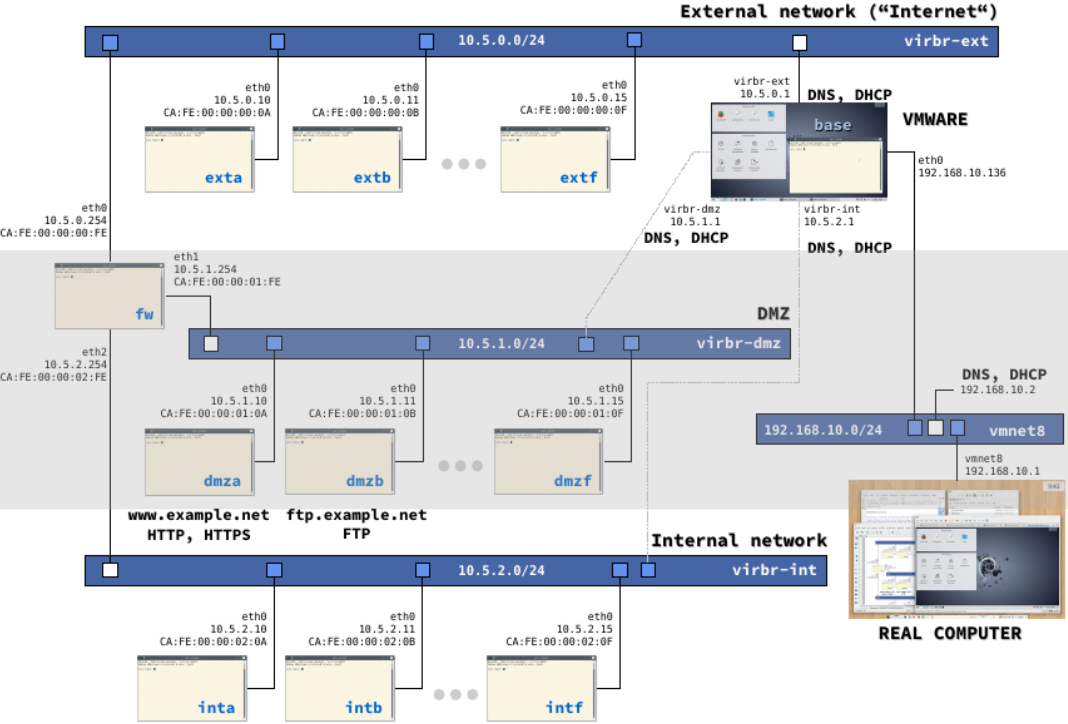}
	\caption{Example infrastructure for evaluation of the Automated Enterprise Architecture Model Mining. \cite{Perez2016}}
	\label{fig:evalnetwork}
\end{figure}

In order to demonstrate the transformation, the prototype is developed with the Flask framework, PM4Py, NetworkML, and Python.
The project is piloted in a test infrastructure with real and synthetic data.
Figure~\ref{fig:evalnetwork} shows a virtual computer network freely available from the \textit{University of Valencia}, called NETinVM~\cite{Perez2016}.
They provide a complete computer network based on virtual images.
The used infrastructure is divided in: internal network, demilitarised zone, and external network.
The internal network has a SSH service.
Machines in the demilitarised zone provide a web service via HTTP, HTTPS and a FTP service.
Furthermore, we used real life event log information for the process miner form the \textit{Eindhoven University of Technology}~\cite{Leoni2020}.

The integration of the new approach into an existing infrastructure must be secured accordingly.
Especially when obtaining the permanent possibilities of automated model generation on request, the accesses have to be secured accordingly with access controls and encrypted data transmission.
Fortunately, a read-only access is required, so access can be limited.
In this way, the concept can be used securely and permanently for an organization in consultation with the IT department.

\begin{figure}[thb]
	\centering
	\includegraphics[width=0.49\textwidth]{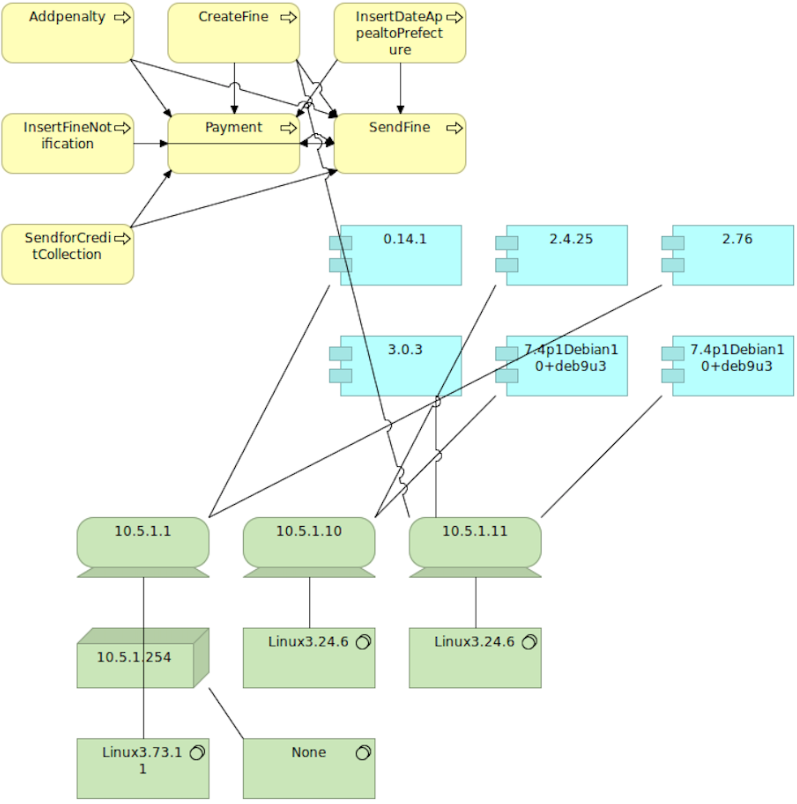}
	\caption{Result of the prototypic implementation.}
	\label{fig:evalresult}
\end{figure}

The result of the prototype is presented in Figure~\ref{fig:evalresult}, where an automatic layout is used.
The evaluation shows that the presented approach fulfils the objective of an \textit{Automatic Enterprise Architecture Model Mining}.
Many EA elements are collected automatically from the computer and the network infrastructure.
The modelling is done automatically with viewpoints according to the specific needs of the stakeholders in an organisation.
Nevertheless, our current prototype shows weaknesses in the automation approach.
Especially the visual representation is still lacking the support of practical solutions of third parties.


The advantage of our solution consists mainly of two complexes of topics.
The first complex is the description in the form of models as a strategic basis for decision-making.
This applies to the updating of legacy systems, harmonization of the IT landscape as well as the future development.
In addition, the clear documentation in models enables communication across levels.
Starting at the lowest level with the description of technical systems, the interaction can be shown in detail.
At the highest level, the information is abstracted and condensed.
Thus, both the interests of the technicians at the lowest level and the requirements of the decision makers at the highest level are fulfilled.
The second main advantage is the fast recognition of potential points of attack.
By identifying new cyber attack opportunities, it is easy to see whether your own IT landscape is also at risk for the attack pattern.
In addition, layered security architectures can be identified and built with the models.
Especially with regard to multi-layer security~\cite{Vuorisalo2018}, there are thus simple control and defence options.
This form of passively oriented vulnerability analysis is a novelty.

\section{Summary}
The need for this research is due to the simplification, acceleration and cost reduction for EA modelling. 
The presented approach is the first holistic way obtaining an automatically enterprise architecture model.
Based on an automated data collection, our engine is modelling generic elements according to a enterprise architecture fragments.
Afterwards, these are transformed to a specific meta-language with respect to the desired EA framework.
As a result, the user receives a holistic EA model of an organization.
In addition, an analysis of the current structure is provided.
EA management can use the data in order to obtain a more detailed and actual picture of the organisation. 
These documents serve for the further development of the company with established methods and reference designs.
It provides possibilities of strategic business decisions.
This gives a company a competitive advantage in the face of ever shorter innovation cycles in the IT sector.
Especially, the visualization support in the field of proactive protective mechanisms and the defence against cyber attacks.

This feasibility analysis is only a first step in the automation of a holistic EA model.
In future developments, the prototype will be expanded with more data collection techniques.
Also, an API integration as extension for \textit{monitoring tools} as well as \textit{security information and event management} is possible.
The export has to be expanded to other EA frameworks as well as EA tools.
Furthermore, the technology for automatic layout and visualization must be expanded.
Last but not least, the reporting can be enhanced with architectural building block from of modular reference blocks within the context of system of systems.
Especially, a cross-layer control and design space exploration is advantageous.

\section*{Acknowledgment}
Special thanks goes to Daniel Rosenstock for the development of the basic prototype~\cite{Rosenstock2020} and his passionate support for this research.

\bibliographystyle{unsrtnat}
\bibliography{IEEEexample} 

\end{document}